# Impact of Surface Charging on Catalytic Processes

Kristof M. Bal,[1] Stijn Huygh, and Erik C. Neyts

Department of Chemistry, University of Antwerp, Universiteitsplein 1, 2610 Antwerp, Belgium

Although significant insights have been obtained into chemical and physical properties that govern to the performance of catalysts in traditional thermal processes, the work on electro-, photo-, or plasma-catalytic approaches has been comparatively limited. The effect of (local) surface charges in these processes, while most likely a crucial factor of their activity, has not been well-characterized and is difficult to study in a consistent, isolated manner. Even theoretical calculations, which have traditionally allowed for the untangling of the atomic-level mechanisms underpinning the catalytic process, cannot be readily applied to this class of problems because of their inability to properly treat systems carrying a net charge. Here, we report on a new, generic, and practical approach to deal with charged semiperiodic systems in density functional calculations, which can be readily applied to problems across surface science. Using this method, we investigate the effect of a negative catalyst surface charge on $CO_2$ activation by supported $M/Al_2O_3$ (M = Ti, Ni, Cu) single atom catalysts. The presence of an excess electron dramatically improves the reductive power of the catalyst, strongly promoting the splitting of $CO_2$ to CO and oxygen. The relative activity of the investigated transition metals is also changed upon charging, suggesting that controlled surface charging is a powerful additional parameter to tune catalyst activity and selectivity.

---

[1] Corresponding author, e-mail: kristof.bal@uantwerpen.be

INTRODUCTION

Efficient and selective heterogeneous catalysts are of enormous technological importance, and will only become more relevant in the context of the rise of new strategies to reduce the emission of greenhouse gasses, requiring novel catalytic processes for the capture and utilization of carbon dioxide through its conversion into value-added chemicals.[1-3] The complexity of the catalytic process and its massive number of chemical and physical degrees of freedom make the design of new catalysts with the right activity and selectivity a daunting task. Untangling all the influencing factors is therefore a crucial step towards a fundamental understanding of the catalytic process, allowing for a more focused optimization strategy. The use of detailed atomistic calculations is an approach that is uniquely suited for a "bottom-up" study of the chemical building blocks that make up the overall catalytic process.[4]

For $CO_2$ activation, energetic and kinetic parameters have been extensively characterized using density functional theory (DFT) calculations on a variety of simple catalyst models, such as flat transition metal surfaces[5-12] and oxide single crystals,[13-21] so that general trends with respect to the chemical properties of these materials can be extracted. With increasing model complexity, however, computational work becomes scarcer, although initial steps have been taken to, as a next phase, investigate the properties of oxide-supported transition metal clusters, as an extension of the work on "pure" materials.[22-25] It is found that the catalyst/support interface plays a significant role in the catalytic activity of the metal, and that reaction mechanisms on a supported cluster can be quite different from those on a pure metal catalyst.[26] The as such obtained insight from incrementally more complex models highlights the power of computational approaches to catalytic $CO_2$ activation.

The properties of a catalyst are dictated by its electronic structure. Besides its primary dependence on the catalyst's chemical and morphological composition as discussed above, it can also be modified through charging, which is an effect at play in electro-, photo- and plasma catalysis. While DFT-based modeling is suited *par excellence* for a controlled investigation of this kind of electronic structure changes, these investigations are severely complicated by the periodicity of the system, in which net charges are impossible to treat straightforwardly.[27] Perhaps for this reason, it is the study of charged surfaces in electrocatalysis for which the most successful modeling approaches have been devised. In this kind of processes, a charged surface can be treated in a neutral simulation cell because the charge distribution in the system naturally follows from its chemical composition; the electrochemical double layer can be explicitly constructed by adding ions to water layers in contact with the surface, self-consistently resulting in a catalyst carrying the opposite charge if box neutrality is enforced.[28,29]

Models of photo- and plasma catalysis, in contrast, require isolated charged catalysts and therefore, in principle, charged simulation cells. In the first application, a small charged slab model is a stand-in for either a localized hole or mobile electron in an excited photocatalyst,[30,31] while in the case of plasma catalysis (i.e., a catalyst surface exposed to a plasma[32]), the catalyst surface in contact with a plasma accumulates a negative charge due to the higher mobility of electrons than plasma ions. In models of photocatalysis, the global charge of the considered surface is not neutral, because the oppositely charged hole or electron is assumed to be localized at a distance beyond the cell size, just like the neutralizing double layer of a plasma sheath is too thick to be explicitly considered in an atomistic model. Especially plasma catalysis remains poorly understood: while physical models[33] and experiments[34] suggest that surface charges can be quite substantial and long-lived, little to nothing is known about its effect on the surface chemistry in a plasma-catalytic system, although this peculiar phenomenon could explain so-called synergistic effects, i.e., the conversion, yield or selectivity is observed to be greater than the sum of pure plasma processing of the gas and pure thermal catalysis.[35-37] Since no direct experimental work in this direction has been carried out, and a controlled set-up to isolate the surface charge effect is difficult to achieve, computational studies become indispensable. Regardless, computational approaches to this type of charged surfaces are limited. Some studies of $CO_2$ adsorption on charged surfaces either attempted to circumvent this limitation by introducing additional electrons from, e.g., adsorbed hydrogen atoms as an approximation of a true surface charge,[30,31] or by ignoring the problem altogether.[38]

In this work, the effect of surface charging in heterogeneous catalysis is explicitly investigated for the first time. A new practical methodology to account for a charged periodic surface in DFT calculations is presented and applied to $CO_2$ activation on a negatively charged supported metal catalyst. As model system, atomically dispersed Ti, Ni and Cu-based transition metal catalysts on the $\gamma$-$Al_2O_3$ (110) surface are considered in order to (1) characterize the structure of single atom catalysts on $Al_2O_3$ and (2) investigate the $CO_2$ reduction ability of these catalysts and the dependence of their chemical properties on the nature of the metal. Besides being a very promising class of materials,[39,40] single atom catalysts also allow us to "purify" the model from the structural complexity of larger supported clusters, models of which require somewhat arbitrary choices of cluster size, structure and orientation.[22-25] We show that the presence of excess electrons in oxide-supported transition metal catalysts dramatically enhances their reductive ability, exemplified by strongly shifting the thermodynamic balance towards $CO_2$ dissociation. These results suggest that controlled charging of the catalyst surface could greatly enhance the efficiency of the $CO_2$ reduction process.

RESULTS AND DISCUSSION

**Practical Treatment of Charged Systems**

Net charges in a (partially) periodic cell are an ill-defined problem because the electrostatic energy of a periodically repeated charged system diverges. Traditional Ewald summation methods avoid this divergence catastrophe by mathematically treating charged systems as if they are immersed in a neutralizing background jellium.[27] This approach causes difficulties when computing energy differences between differently charged systems, as they are also subject to a different artificial background density. While a uniform neutralizing background charge will have little impact on the properties of a spatially homogeneous bulk system, it can induce artifacts when applied to heterogeneous systems or slabs. It furthermore impacts the computation of adsorption energies on a charged surface: since the reference energy of a neutral adsorbate is calculated in a neutral simulation cell, adsorption on a charged slab also includes an artificial contribution from the adsorbate being immersed in the jellium. Energies become dependent on the size of the vacuum region in the simulation cell, and converge poorly with increasing cell sizes, as shown in Supplementary Figure 1.

In this work we propose a simple solution to this problem: by strictly enforcing the neutrality of the box, all unphysical effects from a neutralizing background are avoided. In our approach, global neutrality is achieved through addition of a gas-phase counterion: since slabs carrying a single negative charge were considered in this work, addition of a proton leads to charge cancellation. It should be noted that if this approach were attempted in a plane wave DFT code, charge transfer would occur to the point charge due to use of a non-localized basis set, making it impossible to control the charge of the slab, unless more complicated and less general schemes are adopted.[41] In the CP2K program used here, however, the Kohn-Sham orbitals are expanded in an *atom-centered* (localized) basis, which means that electrons can be made to exclusively localize in the slab if no basis functions are added on the counterion. This means that, if a basis set-free hydrogen atom is placed in the gas phase of a neutral cell, its electron is forced to localize in the catalyst surface. The energy contribution from the dummy point charge is then consistently subtracted out when adsorption energies are computed in the charged cell. A more detailed description of the methodology and an investigation of the effect of the countercharge on computed adsorption energies are given in Supplementary Text 1. The method is in principle readily usable in any DFT code that uses localized basis sets.

In the setup adopted in this work, a single additional electron is considered, which for the surface model used translates to an electron density of $3.68 \times 10^{17}$ m$^{-2}$ or a surface charge density of about $0.06$ C m$^{-2}$. Recent measurements[34] on alumina exposed to a dielectric barrier discharge (DBD) put the plasma-induced surface electron density in the order of $10^{15}$–$10^{17}$ m$^{-2}$, close to values used here. In view of these results, and assuming that the charge penetration depth is no more than 1 nm,[34] the relatively small surface model employed in this work is in fact a realistic approximation of a charged plasma-exposed alumina surface.

**Transition Metal Atom Adsorption on the Al$_2$O$_3$ Support**

For the Ti, Ni, and Cu atoms, different adsorption sites were probed on both the dry and hydrated surface. As discussed in Supplementary Text 2, additional coordination by adsorbed water has an impact on the adsorption characteristics and relative energetics of the surface sites. However, for all metal/surface combinations, the adsorption configuration in which the metal atom is coordinated by two O$_2$ atoms (Figure 1a) was found to be the most favorable, and is the only one considered in the following (all configurations and their energies are given in Supplementary Figures 2 and 3). The effect of surface hydration (and additional OH coordination, Figure 1b) on the metal adsorption energy is limited ($< 10$ %), indicating that transition metal bonding at the surface does not depend strongly on the precise hydration degree or pattern. In all configurations and on all surfaces, Ti adsorbs much more strongly on the oxide surface than Ni or Cu, as depicted in Figure 1c.

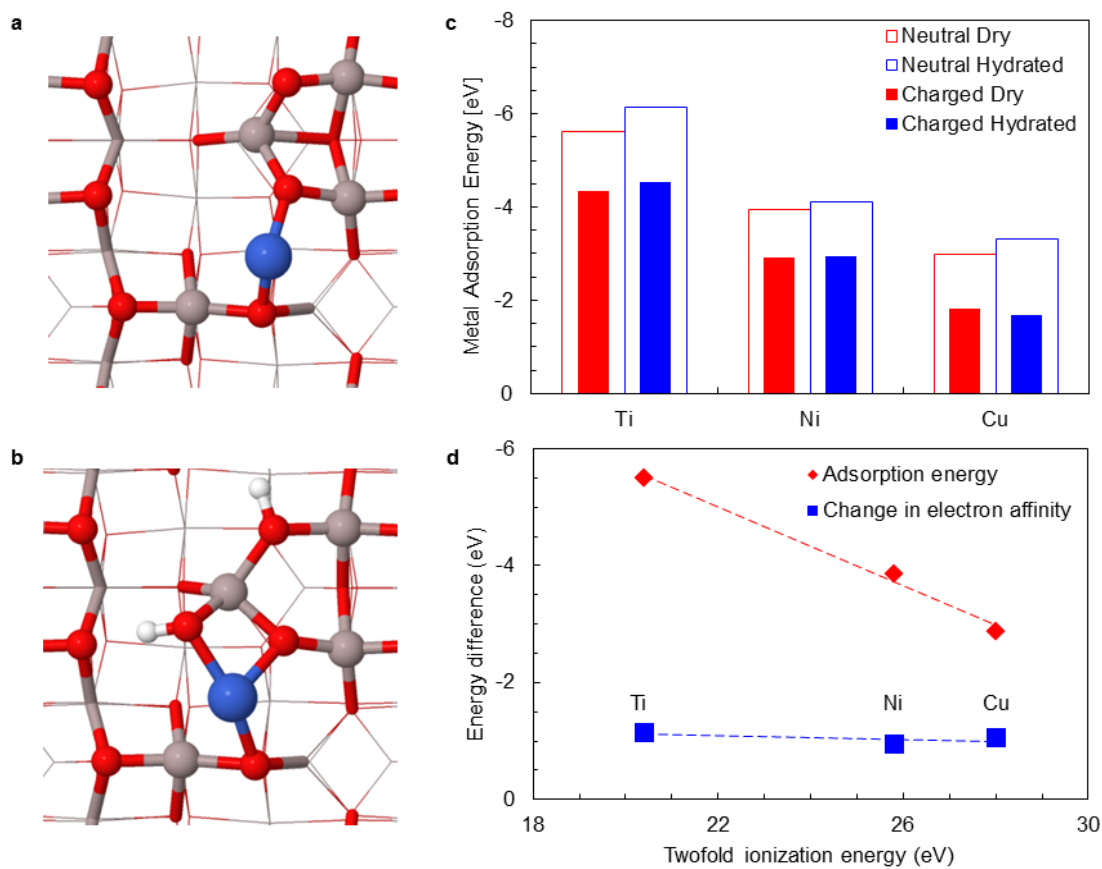

Figure 1: Transition metal adsorption on neutral and negatively charged alumina surfaces. (a) and (b) top view of the most favorable transition metal adsorption configuration on the dry and hydrated surfaces, respectively. Hydrogen: white, oxygen: red, aluminium: gray, and metal: blue. (c) Metal adsorption energies on the two surfaces, with and without extra charge. (d) Correlation of metal binding energies and the change of surface electron affinity $\Delta\chi = E_{ads}(M, neutral) - E_{ads}(M, charged)$ induced by metal binding with metal ionization energies.

Transition metal adsorption on the negatively charged surface is not as favorable. The structures of all metal/support combinations were reoptimized with an additional electron, and absolute metal adsorption energies are about 1 eV smaller in all cases or, alternatively, the electron affinity of the support consistently decreases by this quantity when a transition metal atom is adsorbed. In support of the latter phrasing are the two major indications that the metal/support interaction is mostly ionic in character, with the metal atom adsorbed in its $M^{2+}$ state. First, only very limited mixing of the metal and support electronic states is observed in the projected density of states (PDOS, see also Supplementary Figure 4), which can be associated with a primarily ionic bond. Second, the adsorption energies of the metal atoms on the dry support correlate very well with their combined first and second ionization energies, i.e., the energetic cost of M → $M^{2+}$ + $2e^-$ in the gas phase (Figure 1d). Combined with the near-constant ~1 eV metal-induced downward shift of the support's electron affinity, it can be inferred that metal atom adsorption on $Al_2O_3$ is a redox reaction wherein the support is reduced, which therefore becomes more resistant to further reduction through the absorption of (plasma-supplied) electrons. This reduction of the support upon metal adsorption is of the same magnitude independent of the metal, which is always oxidized to $M^{2+}$ (in this particular configuration), meaning that the support's electron affinity is also modified in the same constant fashion.

**$CO_2$ Adsorption**

$CO_2$ can either chemisorb on the metal atom, or on the $Al_2O_3$ support. In all cases, the adsorbed $CO_2$ molecule adopts a bent carbonate-like structure, with the O−C−O angle deformed by over 40°, as shown on Figure 2a-b.

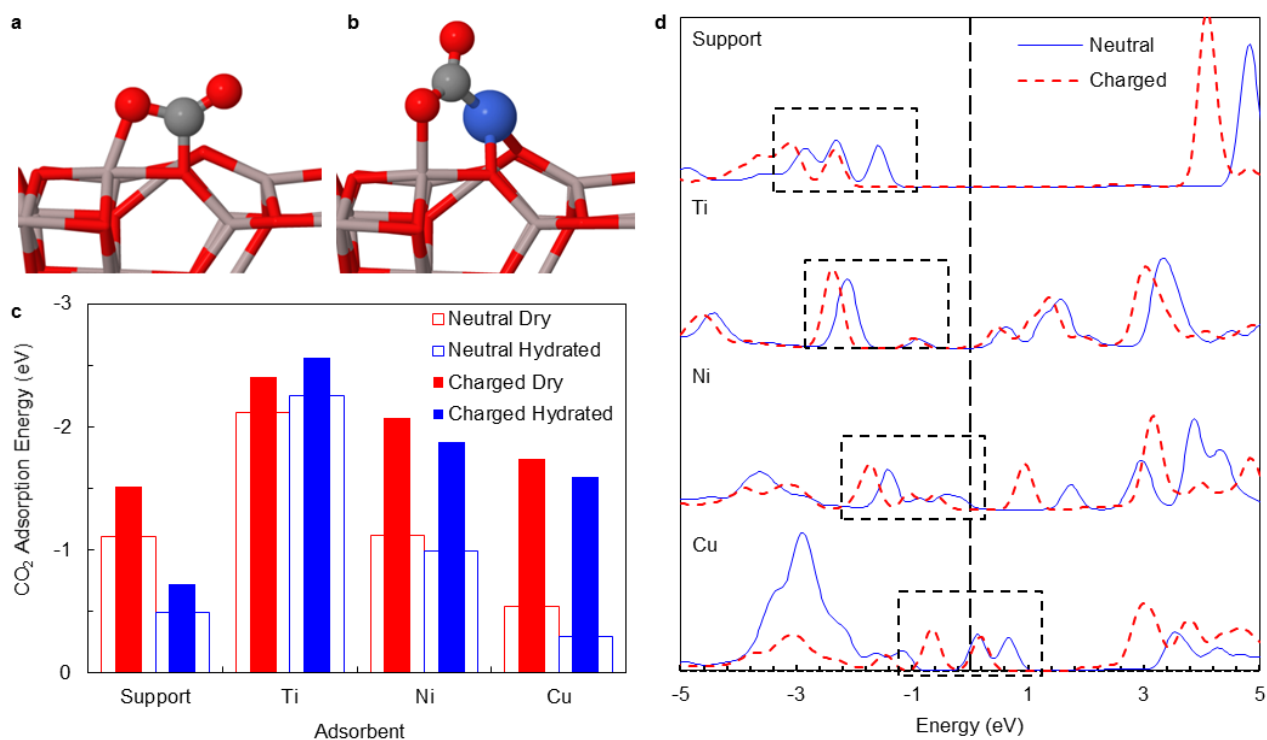

Figure 2: Effect of surface charging on $CO_2$ adsorption. (a) and (b) Most favorable adsorption configuration on the support and supported transition metal atom. (c) Adsorption energies on all sites, with and without extra charge. (d) PDOS of C in $CO_2$ adsorbed on all relevant sites on the dry support, centered on the Fermi level (or, rather, the energy of the highest occupied orbital). The relevant high-lying bonding orbitals are marked with dashed boxes.

On the support, the preferential adsorption site is on an $Al_{IV}$-$O_2$ Lewis pair, forming Al−O and O−C bonds (Figure 2a). Another configuration involving an $Al_{III}$-$O_2$-$Al_{IV}$ site is 0.46 eV less favorable due to the higher Lewis acidity of the $Al_{III}$ site. Indeed, $CO_2$ is a Lewis acid and consequently its affinity with a surface site is proportional with the site's basicity, which is why it is typically used as probe molecule to determine surface basicity. In line with this reasoning, the Lewis acidity of the most favorable $Al_{IV}$ site increases upon hydroxylation of $Al_{III}$,[42] correlating with the lower (by 0.62 eV) $CO_2$ adsorption energy on the hydrated surface. A negative charge transfer, respectively −0.33$e$ and −0.31$e$ on the dry and the hydrated surface, furthermore confirms the Lewis acidic behavior of the $CO_2$ molecule. $CO_2$ chemisorption on the γ-$Al_2O_3$ (110) surface is generally quite similar to adsorption on many other oxides, with adsorption energies in the range of −0.5 to −2.5 eV, formation of a surface carbonate with Lewis basic surface oxygens, strongly bent bi- or tridentate adsorption configurations, and negative charge transfer to the molecule.[13-21] The fairly strong adsorption of $CO_2$ on the alumina support might also increase the retention time of the molecule near the surface, giving it more time to reach an active catalyst site, although it could also increase the competition between metal and support sites.

For all metal/surface combinations, the IVa adsorption configuration is the most stable, and is therefore used in the $CO_2$ adsorption calculations. In all cases, $CO_2$ is found to adsorb in a bridged structure on both the metal atom and the neighboring $Al_{IV}$ surface atom, highlighting the important effect of the support material on the chemical properties of the adsorbed transition metal (Figure 2b). Similar binding modes were observed for larger supported metal clusters, for which the metal/support interface was also the preferred $CO_2$ adsorption location.[22,23] Ni and Cu exclusively bind the $CO_2$ carbon atom, whereas the surface Al atom binds one of its oxygen atoms. Ti, on the other hand, forms an $\eta^2$ complex with the molecule, coordinating both atoms of a C−O bond, while the Al surface atom coordinates the other C−O bond. The ability of the metal/support interface to provide Lewis acid/base pairs is an important property of oxide-supported metal catalysts that can significantly impact its reactivity, with the support material playing in active role beyond merely acting as support for the metal catalyst.

The supported metal atoms show a very diverse $CO_2$ binding behavior, with Ti having the strongest interaction of −2.12 eV (−2.25 eV on the hydrated surface), Ni half as strong with −1.11 eV (−0.99 eV), and Cu even weaker with only −0.54 eV (−0.30 eV), following trends that were established earlier for fcc (100) metal surfaces.[8] In fact, the van der Waals component contributes to about half of the $Cu/CO_2$ interaction (amounting to 0.22 eV and 0.18 eV on the dry and hydrated surface, respectively), pointing to only very limited chemical bonding, insomuch that adsorption on the alumina support is favored over adsorption on the Cu atom. On the dry surface, this is also true for Ni, although hydration greatly diminishes the support's $CO_2$ adsorption ability and favors adsorption on supported Ti or Ni (at least for the particular hydration pattern employed here).

Introduction of an additional electron has a dramatic impact on the adsorption properties, improving the binding characteristics of all $CO_2$ adsorption modes. The magnitude of the effect is the most striking in the case of Cu, which (on the hydrated surface) sees a four-fold increase of the binding energy upon charging, even becoming competitive to Ni. In general, surface charging appears to somewhat "level out" the differences between the metal catalysts, because the effect is much weaker for Ti, which already shows very strong binding with neutral charge.

From a Lewis acid/base theory perspective, negatively charging the surface will naturally increase its basicity and hence improve the binding with the acidic $CO_2$ molecule. To explain the differences between the adsorption modes, their electronic structure must however be analyzed. In particular, examination of the bonding states in the PDOS, and their position relative to the Fermi level is useful here. The comparatively minor surface charging effect on adsorption on the dry support can be attributed by the fact that highest bonding state, formed by overlap of $CO_2$ antibonding π* orbitals with surface $p$ or $d$ states, is fairly low-lying, centered around −2.55 eV (relative to the Fermi level) and shifting to −3.31 eV upon charging; similar observations can be made for $CO_2$ adsorption on supported Ti (−2.13 eV dropping to −2.38 eV). In contrast, the bonding M−$CO_2$ states of the neutral dry Ni and Cu-based catalysts lie partially above the Fermi level, especially explaining the very limited Cu−$CO_2$ bonding: the lower the energy of the metal $d$ states, the more difficult they overlap with the high-lying $CO_2$ antibonding π* orbitals, resulting in a higher energy (i.e., less stabilization) of the bonding states. Surface charging can therefore have a much larger impact in these cases, lowering the bonding states from −1.00 to −1.43 eV (Ni), and 0.46 to −0.28 eV (Cu), relative to the energy of the highest occupied orbital. The relative lowering of the bonding states upon charging is also reflected by the charge of the adsorbed $CO_2$ molecule: increased occupation of these orbitals, which are partially localized on the molecule, leads to a larger electron density (Table 1). In agreement with the PDOS, $CO_2$ bound on Cu is affected the most by surface charging because of the largest shift in energy of the relevant bonding states.

Table 1: Correlation between improved $CO_2$ binding ($\Delta E_{ads}$, in eV) and increased electron density on the adsorbed molecule ($\Delta q$, in elemental charge units).

| Surface | Metal | $\Delta E_{ads}$ | $\Delta q$ |
|---|---|---|---|
| Dry | Ti | −0.29 | −0.15 |
| | Ni | −0.96 | −0.15 |
| | Cu | −1.20 | −0.18 |
| Hydrated | Ti | −0.31 | −0.08 |
| | Ni | −0.89 | −0.12 |
| | Cu | −1.30 | −0.27 |

**Adsorption of Other Molecules on the Support**

While the Lewis acidic $CO_2$ shows improved adsorption behavior on a negatively charged substrate, this is not necessarily a good indicator for molecular adsorption in a general sense. Therefore, we calculated the adsorption energies of water, methane, and carbon monoxide on both the neutral and the charged surface, summarized in Table 2. For water, the hydration energy is considered, i.e., the reaction energy of forming the hydrated surface model from the dry surface. Similarly, for methane, dissociative adsorption into $CH_3$ and H is the studied process. CO is commonly used as basic probe molecule to assess the Lewis acidity of a surface, and is also a major reaction product in $CO_2$ reduction.

Table 2: Influence of surface charging on molecular adsorption energies (eV) at various sites on the $Al_2O_3$ support (d: dry surface, h: hydrated surface).

| Molecule | Site | Neutral | Charged |
|---|---|---|---|
| $H_2O$ | d-$Al_{III}$ | −2.42 | −2.56 |
| $CH_4$ | d-$Al_{IV}$ | −0.38 | −0.46 |
|  | h-$Al_{IV}$ | 0.14 | 0.11 |
| CO | d-$Al_{III}$ | −1.38 | −1.20 |
|  | d-$Al_{IV}$ | −1.07 | −1.02 |
|  | h-$Al_{IV}$ | −1.29 | −1.14 |

Generally, surface charging improves the adsorption behavior of σ-bonded species ($H_2O$ and $CH_4$), but to a much smaller degree (no more than 0.15 eV) due to absence of unoccupied states close to the Fermi level. CO shows the opposite behavior, consistent with its Lewis basic character, also again showing the relative hydration-induced decrease in basicity for this particular configuration. While CO binds on the surface by donating its lone electron pair on C, a negative surface charge can be donated back to CO by partially filling its antibonding π* orbitals, which become more easily accessible because of the higher energy of the surface electronic states: for example, a CO molecule bound at an $Al_{III}$ surface atom has a total charge of 0.06$e$ on the neutral, and 0.02$e$ on the charged surface. Hence, in all cases unoccupied states close to the Fermi level play a crucial role in determining the charge dependence on adsorption; the precise direction of the effect depends on their (anti)bonding nature.

**Impact on Surface Reactions**

The uncatalyzed gas-phase splitting of $CO_2$ (through, e.g., $CO_2 \rightarrow CO + 0.5\,O_2$) is thermodynamically highly unfavorable. On a suitable catalyst, the reaction $CO_2$ (g) → CO (g) + O (ads) can be made more favorable, having a beneficial impact on the overall rate of, e.g., the dry reforming process. Although a structurally simple atomically dispersed catalyst can ostensibly only take part in a small number of reaction mechanisms (direct C−O splitting in this case), the chemical activity of the support material significantly increases the number of possible $CO_2$ activation pathways. While it is not in the scope of this work to obtain a comprehensive picture of the complete catalytic processes of the considered systems, it is useful to have an initial picture of the most common reactions of adsorbed $CO_2$.

To overcome the severe time scale restrictions of traditional MD simulations (~10 ps for DFT-based MD) we use the metadynamics-based[43] *collective variable-driven hyperdynamics* (CVHD) enhanced sampling method[44,45] In CVHD simulations of $CO_2$ adsorbed on the hydrated Ti-based catalyst, the direct splitting reaction $CO_2$ (ads) → CO (ads) + O (ads) (Figure 3a) could be observed at a temperature of 400 K (which is typically achieved in a DBD plasma) after a simulated time of 4.1 μs. Ni is not found to be active at 400 K within the CVHD time scale (which does not, however, rule out the general possibility of a reaction), but does react at 800 K after 0.14 ns. However, no *direct splitting* is observed in this case, but a rather *proton-mediated* mechanism in which a proton is first transferred to the $CO_2$ molecule from an OH group at the support, leading to instantaneous dissociation into CO and OH (Figure 3b) in a near-concerted fashion. The full pathway as observed in CVHD simulations is shown in Supplementary Figure 5 for both mechanisms.

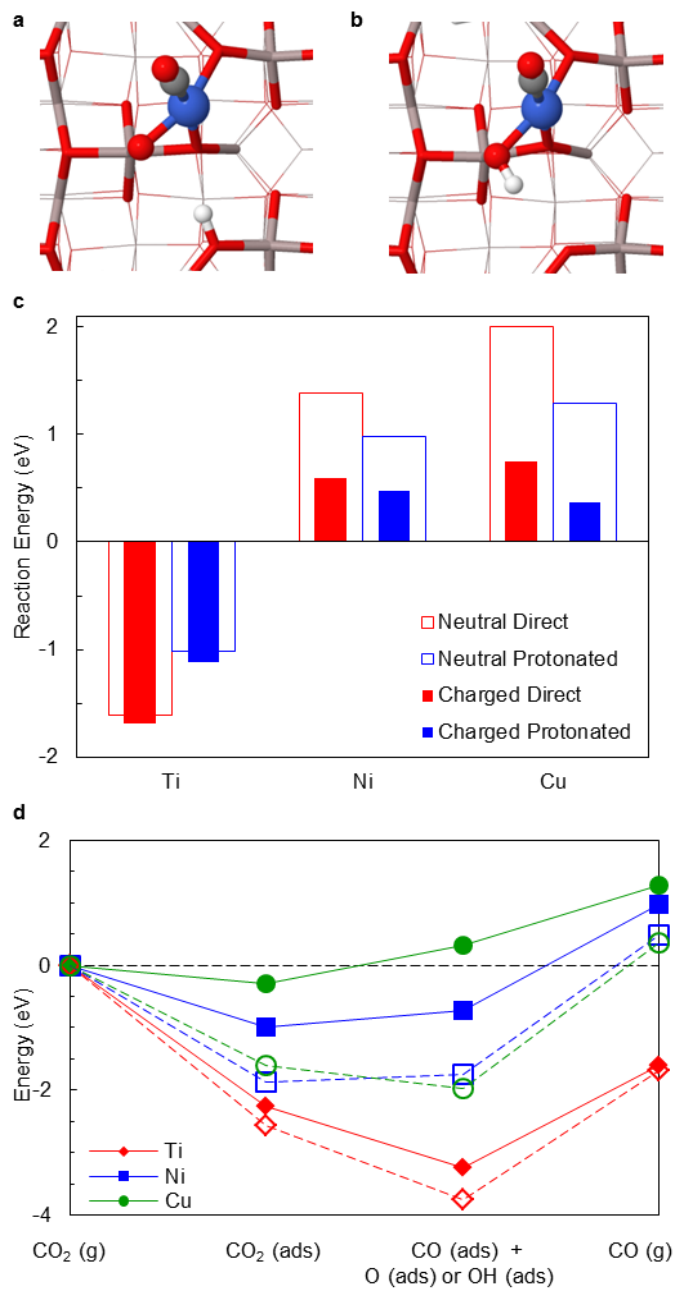

Figure 3: Effect of an excess electron on the reaction energies of $CO_2$ splitting. (a) Product of the direct splitting reaction. (b) Product of proton-mediated splitting. (c) Overall reaction energies for the two studied mechanisms. (d) Most favorable pathways on all metals. Filled symbols and full lines: neutral surface, empty symbols and dashed lines: charged surface.

Motivated by this apparent difference in the reactivity of Ti and Ni, the overall reaction energies of the two competing $CO_2$ activation pathways leading to CO (g) + O (ads) and CO (g) + OH (ads), respectively, were calculated for all metals on both the neutral and charged surface. It is indeed found that Ti is more active towards direct splitting, while Ni and Cu favor a proton-mediated mechanism. Also in agreement with the simulations is the much more favorable reaction energy of initial splitting step on Ti, which reacted at 400 K and exhibits a reaction energy of −0.98 eV, as compared to Ni, which reacted at higher simulated temperatures and has a reaction energy of 0.27 eV.

When examining the effect of an excess electron on the overall splitting process the results largely echo those of the $CO_2$ adsorption, with reactions on Ti relatively unaffected ($\Delta E_{charge}$ = −0.08 eV for direct splitting) and the process on Cu exhibiting a very strong influence ($\Delta E_{charge}$ = −0.92 eV for proton-induced splitting) by the additional negative surface charge (Figure 3c). Interestingly, when decomposing the energetic contributions of the separate process steps (depicted in Figure 3d), it can be seen that the initial $CO_2$ adsorption step is in fact the most affected by the charge, while the subsequent steps are not as dissimilar to their counterparts on the neutral surface. Larger effects are observed again for the desorption of CO, which is more strongly bound on the charged than the neutral surface, in contrast to what was found for the adsorption on the support. This effect, which is unfavorable in the context of CO production, can be attributed to the fact that the CO antibonding π* orbitals can take part in a bonding overlap with metal $d$ states: no antibonding states are therefore occupied even when an excess electron is present (similar as $CO_2$ adsorption), as opposed to the pure σ-bonding on the support. While it becomes more difficult to release CO from the surface upon the charging, the *overall* $CO_2$ splitting process is more favorable. Moreover, CO need not be the final product, but could react further to yield base chemicals such as formaldehyde or methanol, just as well as the additional oxygen atom on the surface can take part in various oxidation processes. This kind of more detailed pathway studies which will be investigated in a future study.

While we have primarily discussed the thermodynamic effects of the excess electron, the kinetics of the catalytic reaction are also of great importance. As a first assessment of the impact of the surface charge on reaction barriers, estimated transition states of the direct splitting reaction were determined. We find that the presence of a negative surface charge consistently lowers the energy of configurations with partially broken bonds such as transition states, lowering the estimated splitting barrier on all metals: from 1.15 eV to 0.75 eV on Ti, 0.80 eV to 0.65 eV on Ni, and 1.26 eV to 0.83 eV on Cu. Through the presence of an additional electron, partially unsaturated atoms in the transition state receive some additional stabilization, hence lowering the apparent reaction barrier and increasing the reaction rate. It must be mentioned that the calculated barriers do not necessarily reflect the lowest energy splitting pathway, but are chosen so as to provide a consistent set of benchmark configurations. For example, in our CVHD simulations we find that $CO_2$ splitting on Ti occurs from a rearranged state in which the molecule is bound exclusively on the metal, as opposed to the metal-support bridge we used as initial state here, as shown in Supplementary Figure 5.

CONCLUSIONS

In the most general sense, electron deposition leads to a chemical reduction of the catalytic surface and, hence, increases its reductive capabilities. Specifically, this phenomenon has a very favorable effect on $CO_2$ activation, with respect to both adsorption strength and overall reaction energy of the splitting reaction. For the strongly oxidizable adsorbed Ti catalyst, this effect is not as pronounced as for Ni and Cu: while all metals formally adsorb in their $M^{2+}$ state, Ti can easily be further oxidized to $Ti^{4+}$, allowing it to act as a strong reducing agent without having to be charged, as evidenced by its strong $CO_2$ activation abilities. The properties of the latter are also largely in line with the redox properties of $TiO_2$ surfaces, resulting from oxygen vacancy creation and annihilation and which allow for efficient reduction of $CO_2$.[14,15]

A less general interpretation of the phenomenon involves viewing the negatively charged catalyst as more Lewis basic, which is appropriate for the description of the bare $Al_2O_3$ support, but is more difficult to apply once adsorbed transition metal clusters have to be considered, as evidenced by the different behavior of CO adsorbed on the support or the metal, respectively. An analysis of the electronic structure of the adsorption complex hence provides the most valuable and robust insight into its response to surface charging.

The major impact of surface charging on the catalytic performance of supported Ni and Cu—even inducing a reversal of their relative activity—demonstrates that conclusions drawn for "conventional" thermal catalysis not necessarily hold for processes involving charged catalysts in, e.g., a plasma. Indeed, the presence of a large surface charge might help explain often-observed but poorly understood synergistic effects in plasma catalysis.

It remains to be seen to what extent the large excess electron-induced effects observed for the systems and reactions of this study are applicable to other catalysts and processes. Different support materials (e.g., semiconductors rather than isolators), larger supported clusters, transition metal surfaces and a more exhaustive set of redox processes should all be considered in order to assess the influence of a negative surface charge on catalysts in a more general sense, in which the methodology outlined in this work can provide the template for such a systematic undertaking. However, the results presented in thus work already point to a phenomenon with potentially far-reaching consequences: by varying the discharge parameters of the plasma and the degree of electron deposition on the plasma-facing catalyst, its Lewis acidity and redox properties can be modified as well. Thus, controlling the electron deposition on a catalyst opens another avenue towards activity and selectivity control of a plasma-catalytic process.

## METHODS

**General Methodology**

All DFT calculations were carried out with the Quickstep module in the CP2K 4.1 package.[46,47] Energies and forces were computed using the Gaussian and plane wave (GPW) method[48] employing Goedecker-Teter-Hutter (GTH) pseudopotentials[49,50] for the core-valence interactions and a polarized double-ζ (m-DZVP) basis set[51] to expand the Kohn-Sham valence orbitals. An auxiliary plane wave basis set defined by a cutoff of 1200 Ry was used to expand the electron density. Exchange and correlation were treated with the PBE functional,[52] supplemented by Grimme's D3 dispersion correction[53] in its Becke-Johnson damping form.[54] Atomic partial charges were calculated by the self-consistent Hirshfeld-I scheme.[55] Molecular adsorption energies were calculated as $E_{ads} = E_{mol+surface} - E_{mol} - E_{surface}$ and are reported without thermal or zero-point corrections. Reaction barriers were estimated with the nudged elastic band method;[56] transition state structures were only optimized for the neutral slabs, and single point calculations were carried out in the presence of an excess electron.

**Surface Model**

Calculations were carried out on a slab of the γ-$Al_2O_3$ structure proposed by Digne *et al.*[57] The (110) surface was modeled as a 2 × 2 supercell containing 240 atoms, corresponding to six layers of which the bottom two were kept fixed at their bulk positions. The simulation cell dimensions were 16.1606 × 16.8106 × 40 Å$^3$, while calculations involving a surface charge employed a larger box with a Z length of 100 Å, with the counter charge placed at a Z position of 40 Å. Periodic boundary conditions were not applied along the Z direction to avoid self-interaction of the slab; calculations involving isolated atoms or molecules were also carried out in these cell sizes. The surface exposes both coordinatively unsaturated Al and O atoms. Tri- ($Al_{III}$) or tetracoordinate Al ($Al_{IV}$) atoms provide Lewis-acidic sites, whereas di- ($O_2$) and tricoordinate ($O_3$) surface atoms are Lewis basic. Although the (110) surface termination is the most common, it is not stable in its "dry" form, which is why a hydrated variant was also considered in this work (structure s1a from ref. 42) containing 4 adsorbed water molecules, corresponding to a density of about 3 OH nm$^{-2}$. This surface is the most stable adsorption configuration of a single adsorbed water molecule per unit cell, which is dissociated into an OH group adsorbed on the $Al_{III}$ site and a proton bonded with an $O_2$ atom. Comparison of the two surfaces allows assessing the impact of adsorbed water on the properties of the $Al_2O_3$ support.

**Cross-Validation Checks**

Unless noted otherwise, the abovementioned PBE-D3 based methodology was employed for all calculations, but a small subset of structures was re-optimized using different exchange-correlation functionals in order to assess the reproducibility or our results and their dependence on the chosen approximations. These additional calculations employed the D3-corrected revPBE[58] and TPSS[59] functionals, the "plain" uncorrected PBE functional and the PBE-rVV10 functional. This latter functional combines PBE exchange-correlation with the nonlocal van der Waals correlation component of the rVV10 functional,[60,61] and was generated in this work by refitting its $b$ parameter[62] against an accurate binding curve of the Ar dimer (see Supplementary Text 3 and Figure 6).[63] A more detailed description of all cross-checks is given in Supplementary Text 4, and Tables 2 and 3. In general, however, we find that the sensitivity of our results on the choice of the density functional approximation is very small, and has therefore no impact on the general conclusions presented here.

**Accelerated Molecular Dynamics Simulations**

In the MD simulations, a reduced plane wave cutoff of 400 or 600 Ry and box $Z$ length of 25 Å was used, with full periodic boundaries. The equations of motion of the Nosé-Hoover chain were integrated with a 0.5 fs time step. Before production runs, each system was equilibrated for 1 ps at the desired temperature. CVHD biasing forces were calculated and applied with the PLUMED plugin.[64] In CVHD, bond distortions were biased up to a maximal value of 0.5 (50 % bond elongation compared to equilibrium) through addition of a repulsive Gaussian of height 0.01 eV and width 0.05 every 10 fs, with a well-tempered bias factor of 20. More details about the choice of CVHD parameters can be found elsewhere.[44,45] The boost factors that were obtained range from ~100 at 800 K, to over $3 \times 10^6$ at 400 K.

ACKNOWLEDGEMENTS

K.M.B. is funded as PhD fellow (aspirant) of the FWO-Flanders (Fund for Scientific Research-Flanders), Grant 11V8915N. The computational resources and services used in this work were provided by the VSC (Flemish Supercomputer Center), funded by the Research Foundation - Flanders (FWO) and the Flemish Government – department EWI

# Supplementary Information for: Impact of surface charging on catalytic processes

## Supplementary Text 1: Treatment of Charged Systems

To assess the influence of the handling of charged slabs on computed properties, we first checked the "straightforward" approach of a slab in a fully periodic cell with a net charge. In Supplementary Figure 1(top), the effect of the cell size (by varying its $Z$ length) on both the total energy of a dry slab, as the adsorption energy of $CO_2$ on it, is shown. Even for very large cells, no clear convergence is observed.

In our countercharge-based approach, a basis set-free hydrogen atom was placed above the slab to introduce the positive dummy countercharge. The accuracy of this approach hinges on the assumption that if the energetic contribution from the point charge is the same for all systems, adsorption energies are not affected because its effect is cancelled when subtracting the energies of the slab + adsorbate and the clean slab. To verify this assumption, energies were computed for different $Z$ positions of the countercharge, depicted in Supplementary Figure 1(bottom), employing a $Z$ box length of 100 Å.

The interaction energy of the countercharge and the slab is linearly dependent on their mutual distance, which is the expected behavior for a point charge interacting with an infinite charged plate. At sufficient separation, the effect of adsorbed species on this interaction becomes negligible, and the computed adsorption energy converges. A $Z$-position higher than 30 Å (or a distance of ~20 Å) suffices, and the value of 40 Å used in our production calculations is a very conservative choice.

## Supplementary Text 2: Transition Metal Adsorption

For the Ti, Ni, and Cu atoms, different adsorption sites were probed on both the dry and hydrated surface. These configurations are based on those of the Al atoms in the "next" surface layer; that is, the location of the Al atoms in a hypothetical additional atomic layer atop the actual surface layer in this work. The sites are named after the Al atom on which they were based meaning that, e.g., the d-III configuration is based on the position of an $Al_{III}$ atom in the hypothetical top layer. The possible metal atom adsorption configurations are depicted in Supplementary Figure 2.

The most stable adsorption site on the dry surface for all three metals is d-IVa, at which two highly unsaturated (dicoordinate) O atoms can coordinate the metal atom, while d-III and d-IVb have a similar stability (Supplementary Figure 3, empty bars). Hydration has a drastic impact as it introduces OH groups on the surface that can provide additional coordination of the metal atom; for Cu, the h-IVa and h-IVb configurations are essentially degenerate (Supplementary Figure 3, filled bars). Remarkably, however, the IVa configuration remains the most favorable in all considered cases. It is conceivable that at higher water coverages or with different hydration patterns, a wider array of adsorption configurations are possible; our calculations provide a first step towards a more complete understanding of transition metal adsorption on realistic alumina surfaces.

De deformation of the surface upon transition metal adsorption is quite limited in case of Ni and Cu, mainly amounting to a small (< 0.1 Å) elongation of the Al–O bonds of the coordinating surface O atoms. Ti, however, can have a much larger impact on the support surface structure: in the d-Ti-IVa configuration, the $Al_{IVb}$-$O_{2b}$ bond is extended from 1.69 Å to 2.41 Å, essentially converting $Al_{IVb}$ into an $Al_{III}$ site. Therefore, Ti can strongly affect the activity of the surface region close to it.

## Supplementary Text 3: Parameterization of a PBE-rVV10 Combination

The rVV10 nonlocal correlation functional,[1] which is a revision of the VV10 functional that is better suited for plane wave calculations, consists of a standard rPW86PBE generalized gradient approximation (GGA) exchange-correlation functional combined with a nonlocal correlation term due to Vydrov and Van Voorhis.[2] Combining a different base GGA functional with the nonlocal correlation term as van der Waals correction can be achieved by refitting the adjustable parameters in the nonlocal term.[3] Of these parameters, $b$ controls the short range behavior, and $C$ the long range. Hence, only $b$ should be refitted when the base functional is changed, and the original value for $C = 0.0093$ can be retained.

It is found that even for the original VV10 implementation, optimal $b$ values depend on the fitting target. When fitting against a standard test set, $b = 5.9$ is obtained, whereas a correct description of noncovalent interactions in liquid water[4] or layered solids[5] requires $b > 9$. We therefore decided to only fit PBE-rVV10 against a simple "fundamental" reference system, namely the argon dimer binding curve.[6] This way, accuracy for specific systems is sacrificed in favor of a more clear and simple parameterization strategy with little empiricism.

We use an accurate estimate of the attractive part of the Ar dimer binding curve as reference.[6] Calculations were carried out in a 25 × 25 × 25 Å³ box using a 1200 Ry cutoff for the density. To minimize the basis set superposition error, the very large aug-QZV3P basis set was used to expand the Kohn-Sham valence orbitals. All calculations were evaluated self-consistently.

The value of $b = 9.5$ was found to be optimal (minimizing the RMS deviation from the reference), and the resulting PBE-rVV10 functional yields a good binding curve for the Ar dimer, depicted in Supplementary Figure 6. Due to the shallowness of the curve, recovering the exact minimum is difficult with PBE-rVV10, but in overall terms, energetics are well-described.

## Supplementary Text 4: Computational Consistency Checks

The stability of the various $CO_2$ adsorption sites was reinvestigated using a set of different approximations of the exchange-correlation energy. This small series of calculations should not be seen as a true "benchmark" of these approximations (due to the unavailability of a reference to measure against) but rather a "consistency check" to verify to what extent our conclusions depend on the computational choices that were made.[7] Specifically, the treatment of dispersion interactions was verified by comparing the "plain" uncorrected PBE functional with its D3- and rVV10-corrected variants, whereas the role of the underlying functional was assessed by comparing three common approximations to the exchange-correlation energy: the widely used PBE GGA functional (and used throughout the rest of this work), its revPBE variant (which tends to perform better for thermochemistry[8] and surface science[9]), and the TPSS meta-GGA (which represents a higher, more advanced, rung on "Jacob's ladder" of density functional approximations[7]). These functionals were applied in their D3-corrected form.

Inspection of Supplementary Table 1 demonstrates that all methods are in close agreement with respect to the relative stability of the $CO_2$ adsorption sites and, for the dispersion corrected methods, even in terms of absolute adsorption energies. Furthermore, all methods give hydration energies in good agreement with the value of –2.34 eV/$H_2O$ calculated by Wischert et al.[10] It should be noted, however, that our plain PBE calculations do not recover the results of Pan et al.[11] of $CO_2$ adsorption on the γ-$Al_2O_3$ (110) surface, as these authors reported a different relative ordering of d-III and d-IVb configurations, with adsorption energies of –0.43 and –0.27 eV, respectively, and a bidentate structure for d-III rather than a tridentate. The origin of this discrepancy is unclear to us, but our results are more in line with those of other oxides and the relative Lewis acidity of the sites.

It can be seen that the effect of including dispersion corrections is quite large, amounting to about 0.3 eV for $CO_2$ adsorption. The necessity of including dispersion was previously demonstrated for $CO_2$ adsorption on $TiO_2$[12] and is confirmed here for $Al_2O_3$. It is reassuring to observe that the D3 and rVV10 methods, although based on different principles (atom pairwise and nonlocal density based, respectively), give very similar results, thus validating each other's applicability to this system. In all other calculations (of the neutral systems), we adopted the PBE-D3 method for all geometry optimizations but, because we deal with the rather challenging case of metal-containing systems, follow Hujo and Grimme's recommendation[3] and verify PBE-D3 results against single point cross-checks with PBE-rVV10. These tests are collected in Supplementary Table 2 and Supplementary Table 3, and show that while rVV10 consistently gives smaller adsorption energies for transition metal atoms (in the order of 0.1–0.2 eV), it is very close to D3 for adsorption of $CO_2$ on those metals and, most importantly, gives the same energy differences between adsorption sites.

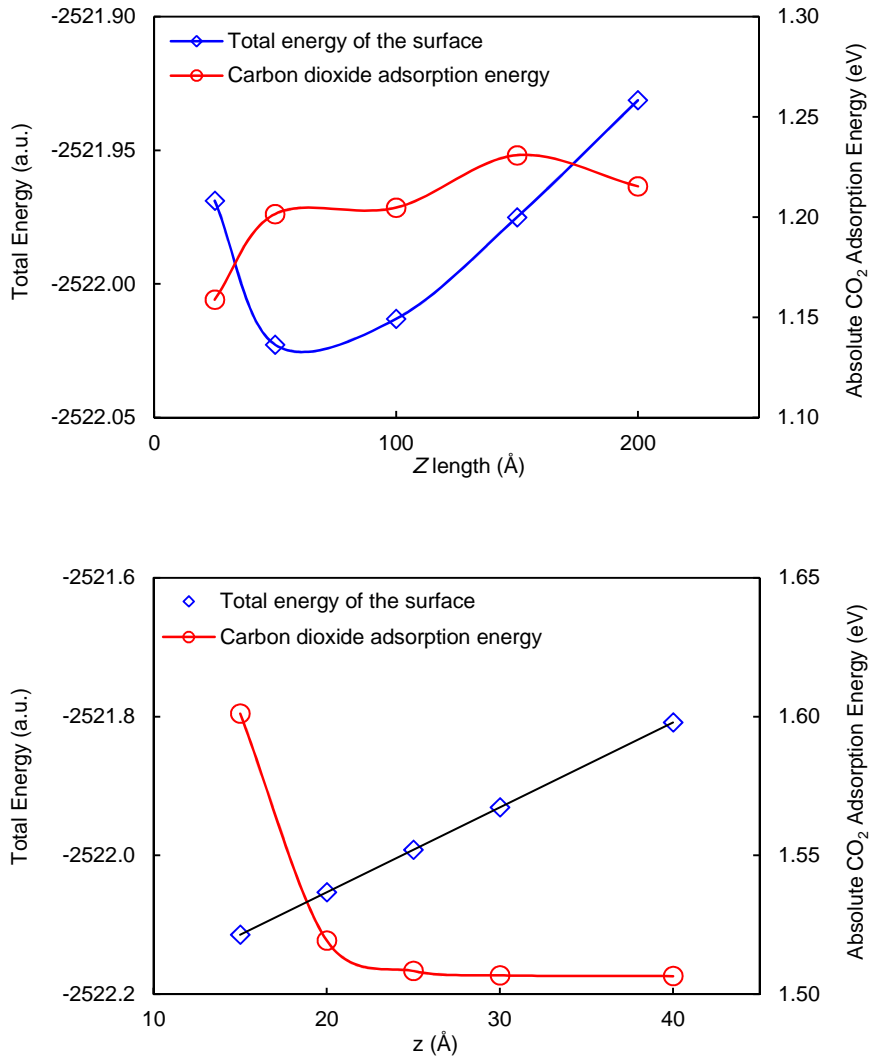

**Supplementary Figure 1 | Convergence of computed $CO_2$ adsorption energies (in the d-IVb configuration) on a negatively charged slab.** (top) Negatively charged slab without neutralizing charge and its dependence of the cell size. (bottom) Negatively charged slab with neutralizing charge and its dependence on the position of the neutralizing countercharge. For the total energies, a straight line is fitted.

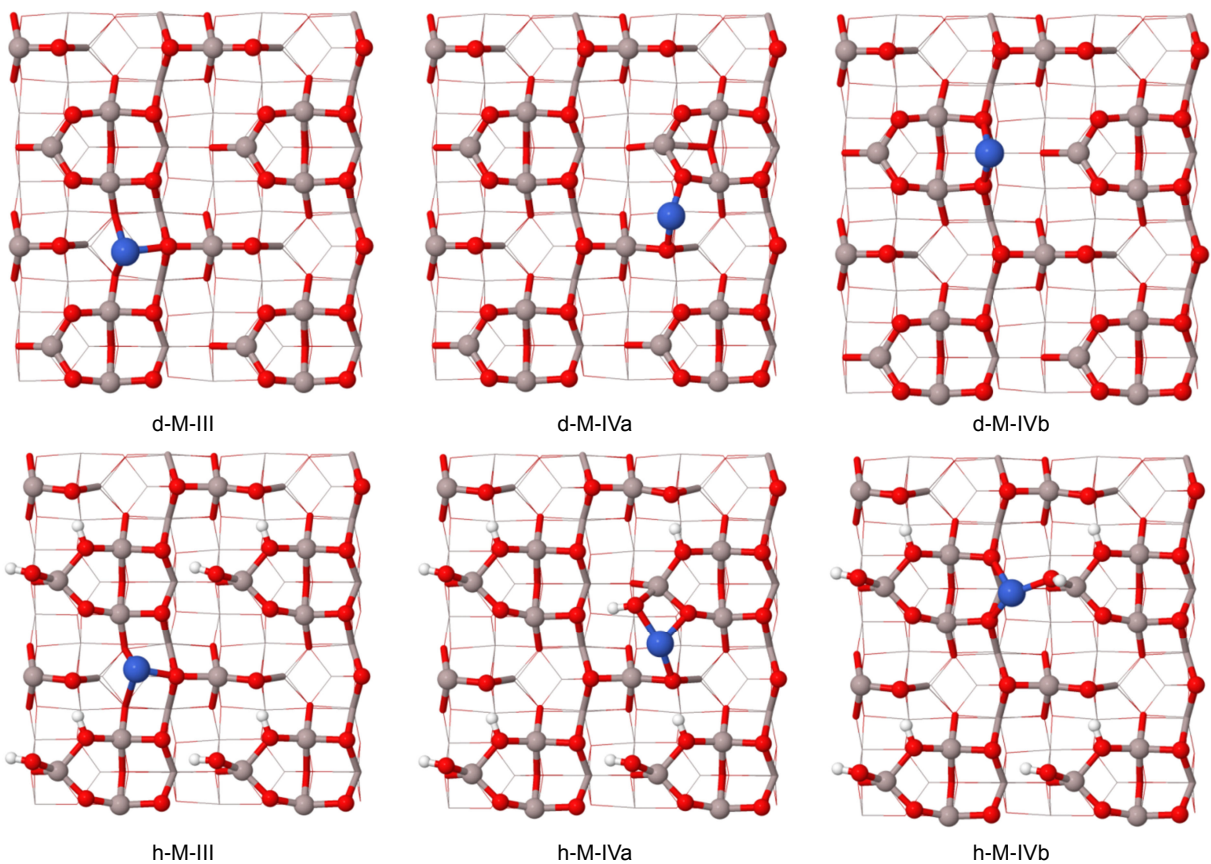

**Supplementary Figure 2 | Metal adsorption configurations on the dry and hydrated surface.** Configurations starting with "d-" denote the dry surface, while those with "h-" describe the hydrated surface. Hydrogen: white, oxygen: red, aluminium: gray, and metal: blue. The depicted configurations are for adsorbed Ni, but do not differ significantly for Ti and Cu.

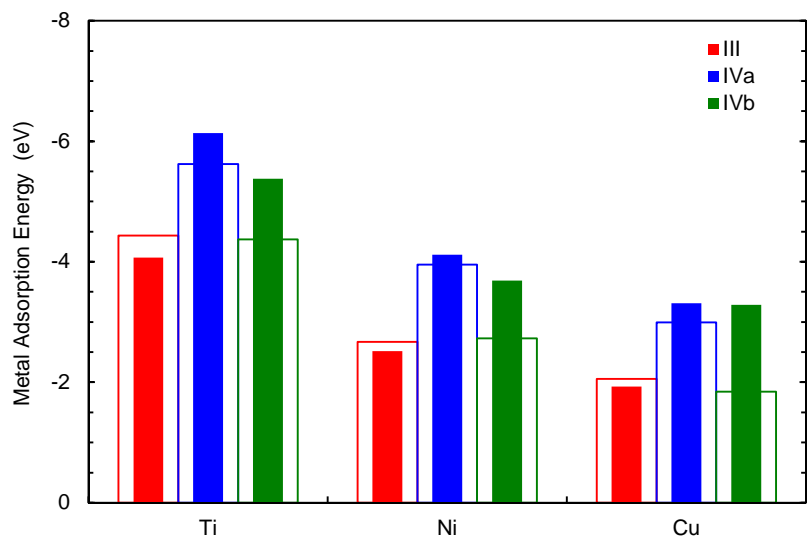

**Supplementary Figure 3 | Relative energies of the metal adsorption configurations on the dry and hydrated surface.** Empty bars reflect adsorption on the dry surface, filled bars represent the hydrated surface.

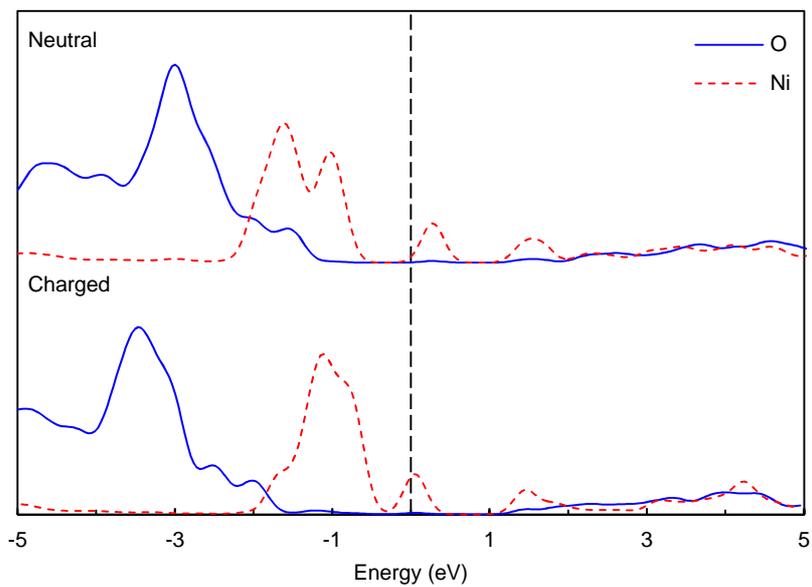

**Supplementary Figure 4 | Projected densities of states (PDOS) for supported Ni.** Shown are the states of Ni and surface oxygens of the dry surface. Energies are centered on the Fermi level. It can be seen that mixing of metal and surface states is essentially nonexistent.

Ti @ 400 K

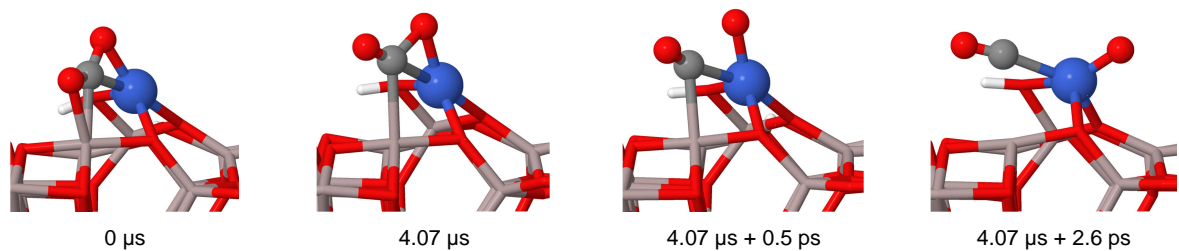

| 0 µs | 4.07 µs | 4.07 µs + 0.5 ps | 4.07 µs + 2.6 ps |

Ni @ 800 K

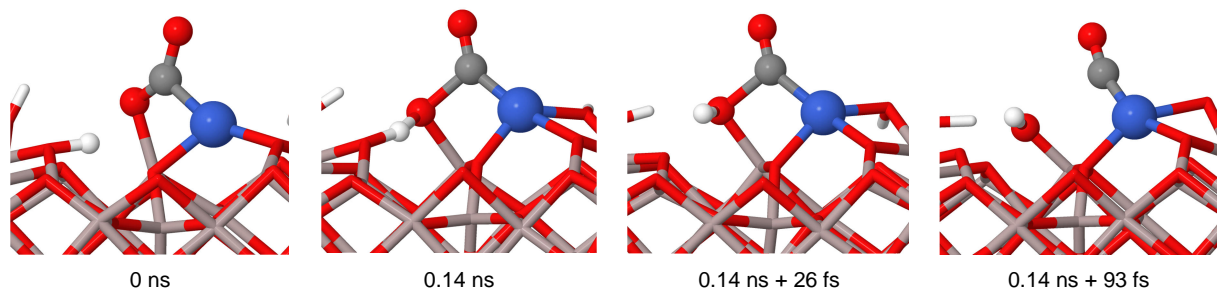

| 0 ns | 0.14 ns | 0.14 ns + 26 fs | 0.14 ns + 93 fs |

**Supplementary Figure 5 | Proton-induced $CO_2$ splitting by supported metal atoms.** Reaction steps observed in CVHD simulations for Ti at 400 K and Ni at 800 K. Accelerated time is given below each frame. Given the time scales, it can be concluded that for Ni, proton transfer and C–O splitting are essentially concerted.

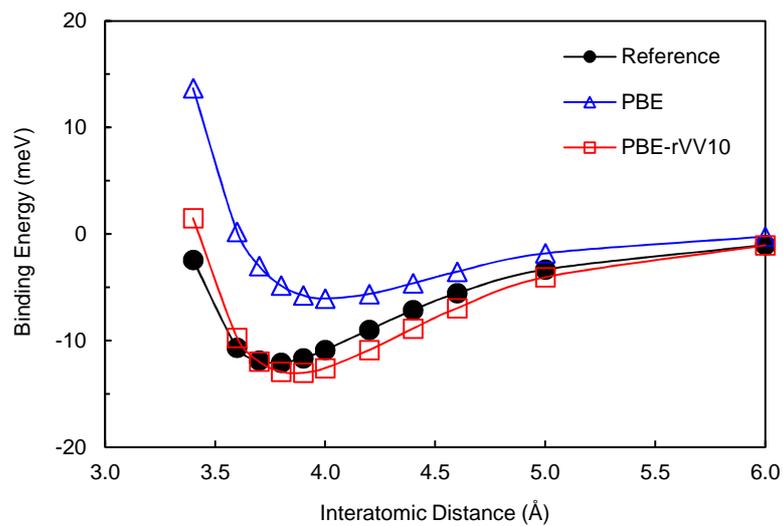

**Supplementary Figure 6 | Ar dimer binding curve.** Energies from calculations employing uncorrected PBE and optimized PBE-rVV10 functionals are compared against an accurate reference.

**Supplementary Table 1 | Computational consistency checks.** Hydration (per $H_2O$ molecule) and $CO_2$ adsorption energies (in eV) for different adsorption configurations, as calculated by different density functional methods.

|  | PBE | PBE-D3 | PBE-rVV10 | revPBE-D3 | TPSS-D3 |
| --- | --- | --- | --- | --- | --- |
| $H_2O$ | −2.23 | −2.42 | −2.36 | −2.36 | −2.42 |
| d-$CO_2$-III | −0.37 | −0.65 | −0.64 | −0.52 | −0.66 |
| d-$CO_2$-IVb | −0.82 | −1.11 | −1.07 | −1.02 | −1.17 |
| h-$CO_2$-IVb | −0.20 | −0.49 | −0.44 | −0.43 | −0.56 |

**Supplementary Table 2 | Computational consistency checks for transition metal adsorption.** Comparison of D3 and rVV10 dispersion corrections for transition metal adsorption on the alumina surface. All calculations use PBE-D3 geometries. Adsorption energies are in eV.

|  | PBE-D3 | PBE-rVV10 |
|---|---|---|
| M @ dry surface | | |
| d-Ti-III | −4.44 | −4.24 |
| d-Ti-IVa | −5.62 | −5.49 |
| d-Ti-IVb | −4.37 | −4.26 |
| d-Ni-III | −2.67 | −2.53 |
| d-Ni-IVa | −3.95 | −3.85 |
| d-Ni-IVb | −2.73 | −2.61 |
| d-Cu-III | −2.05 | −1.90 |
| d-Cu-IVa | −3.00 | −2.87 |
| d-Cu-IVb | −1.84 | −1.71 |
| M @ hydrated surface | | |
| h-Ti-III | −4.07 | −3.88 |
| h-Ti-IVa | −6.14 | −5.97 |
| h-Ti-IVb | −5.38 | −5.22 |
| h-Ni-III | −2.52 | −2.38 |
| h-Ni-IVa | −4.12 | −3.99 |
| h-Ni-IVb | −3.69 | −3.57 |
| h-Cu-III | −1.93 | −1.78 |
| h-Cu-IVa | −3.31 | −3.17 |
| h-Cu-IVb | −3.28 | −3.16 |

**Supplementary Table 3 | Computational consistency checks for CO$_2$ adsorption.** Comparison of D3 and rVV10 dispersion corrections for CO$_2$ adsorption on the alumina surface. All calculations use PBE-D3 geometries. Adsorption energies are in eV.

|                        | PBE-D3 | PBE-rVV10 |
|------------------------|--------|-----------|
| d-CO$_2$- IVb          | −1.11  | −1.06     |
| h-CO$_2$- IVb          | −0.49  | −0.44     |
| d-Ti-CO$_2$            | −2.12  | −2.09     |
| d-Ni-CO$_2$            | −1.11  | −1.09     |
| d-Cu-CO$_2$            | −0.54  | −0.52     |
| h-Ti-CO$_2$            | −2.25  | −2.23     |
| h-Ni-CO$_2$            | −0.99  | −1.01     |
| h-Cu-CO$_2$            | −0.30  | −0.27     |

# Suplementary References

Note: some of these references already appeared in the main manuscript, and are simply repeated here for the reader's convenience.